\documentclass[num-refs]{wiley-article}
\usepackage{siunitx}
\usepackage{placeins} % for \FloatBarrier command

\newcommand{\subf}[2]{%
  {\small\begin{tabular}[t]{@{}c@{}}
  #1\\#2
  \end{tabular}}%
}

\title{Strain measurement by contour analysis}

\author[1,2]{Georg C. Ganzenm\"uller}
\author[2]{Puneeth Jakkula}
\author[1,2]{Stefan Hiermaier}

\affil[1]{Albert-Ludwigs-Universit\"at Freiburg, Institute for Sustainable Systems Engineering, INATECH, Emmy-Noether Str. 2, 79110 Freiburg, Germany}
\affil[2]{Fraunhofer Institute for High-Speed Dynamics, Ernst-Mach-Institut, EMI, Ernst-Zermelo Str. 4, 79104 Freiburg, Germany}

\corraddress{Georg C. Ganzenm\"uller, Albert-Ludwigs-Universit\"at Freiburg, Sustainable Systems Engineering, INATECH, Emmy-Noether Str. 2, 79110 Freiburg, Germany}
\corremail{georg.ganzenmueller@inatech.uni-freiburg.de}

\fundinginfo{Carl-Zeiss Stiftung, \textit{Skalen\"ubergreifende
Charakterisierung robuster funktionaler Materialsysteme}; Gips-Schüle Stiftung, \textit{named chair: Professur für Nachhaltige Ingenieursysteme}}

\runningauthor{G. C. Ganzenm\"uller et al.}

\begin{document}
\begin{frontmatter}
\maketitle

\begin{abstract}
The abstract is located on the next page for formatting reasons.
\end{abstract}
\end{frontmatter}

\pagebreak
\section*{Abstract}
\textbf{Background}:
The determination of yield stress curves for ductile metals from uniaxial material tests is complicated by the presence of tri-axial stress states due to necking. A need exists for a straightforward solution to this problem.\\
\textbf{Objective}:
This work presents a simple solution for this problem specific to axis-symmetric specimens. Equivalent uniaxial true strain and true stress, corrected for triaxiality effects, are calculated without resorting to inverse analysis methods.\\
\textbf{Methods}:
A computer program is presented which takes shadow images from tensile tests, obtained in a backlight configuration. A single camera is sufficient as no stereoscopic effects need to be addressed. The specimen's contours are digitally extracted, and strain is calculated from the contour change. At the same time, stress triaxiality is computed using a novel curvature fitting algorithm.\\
\textbf{Results}:
The method is accurate as comparison with manufactured solutions obtained from Finite Element simulations show. Application to 303 stainless steel specimens at different levels of stress triaxiality show that equivalent uniaxial true stress -- true strain relations are accurately recovered.\\
\textbf{Conclusions}:
The here presented computer program solves a long-standing challenge in a straightforward manner. It is expected to be a useful tool for experimental strain analysis.\\
\\
\textbf{Keywords} strain analysis; optical methods; tensile testing; stress triaxiality

\section{Introduction}

G'Sell \textit{et al.} \cite{gsellVideocontrolledTensileTesting1992} were the
first ones to perform strain analysis for cylindrical specimens using contour
extraction from digital images. We revisit their idea and provide
an estimation of the strain measurement accuracy by comparing against FEM
simulation. The principle of the contour method is to measure the evolution of the diameter and compute the
strain from the diameter change, assuming knowledge of how radial strain is
related to longitudinal strain. This is, of course, trivial for ductile metals
during plastic deformation, as this is typically an isochoric process.

The idea of G'Sell \textit{et al.}, to use a single camera with backlight
illumination to obtain the specimen contours, seems not to have found many
adopters, even though it is easy to use. Digital Image Correlation (DIC)
techniques with their very general capability have replaced specialized
techniques, and the contour analysis method is only applicable to cylindrical
specimens in a straightforward manner. In our opinion, however, contour analysis
provides a very good alternative to DIC in this special case, as it allows for
evaluating the stress triaxiality. In this way, experimental force/displacement
date can be directly converted to \textit{uniaxial} true stress / true strain
curves, suitable as input data for simulations.

A quite sophisticated approach to strain estimation by contour analysis was
taken by Arthington \textit{et al}.
\cite{arthingtonCrosssectionReconstructionUniaxial2009}. They used three
projections of the contours, allowing them to compute also plastic Poisson's
ratio, thus eliminating the need for a-priori knowledge of this quantity.
Although very promising, their approach lacks the simplicity of the original
method and requires a -- relatively -- complex algorithm to post-process the
image data. The aim of the work presented here is to provide theoretical
background and an estimate of the accuracy to expect for the simple algorithm
along with a publicly accessible open-source computer code for other researchers
to use.

The remainder of this work is organized as follows. We begin by describing the
method, and in particular, the salient feature of the computer algorithm used to
evaluate the acquired images. Then, synthetic image data, obtained from FEM
simulations, are used to analyze the accuracy of the method. Subsequently, the
method is applied to tensile tests of a ductile stainless steel. Here, specimens
with both notched and parallel gauge regions are considered, and the effects of
stress triaxiality are quantified and removed from the data to yield uniaxial
yield stress curves. Finally, we summarize our work by pointing out the
strengths and weaknesses.

\section{Description of the Method}

\subsection{Strain and stress measures}

The purpose of this section is to define the stress and strain measures referred
to in this work. We consider the true stress, which is the one-dimensional
projection of the Cauchy stress, i.e., force per current cross section. The
corresponding 1D strain measurement is the logarithmic strain. This pair of
stress and strain measures is applicable for use with most large-deformation
Finite Element codes. Such codes predominantly employ material models defined
using true strain and true stress, assuming that the stress state is 1D.

\begin{figure}
	\centering
	\includegraphics[width=6cm]{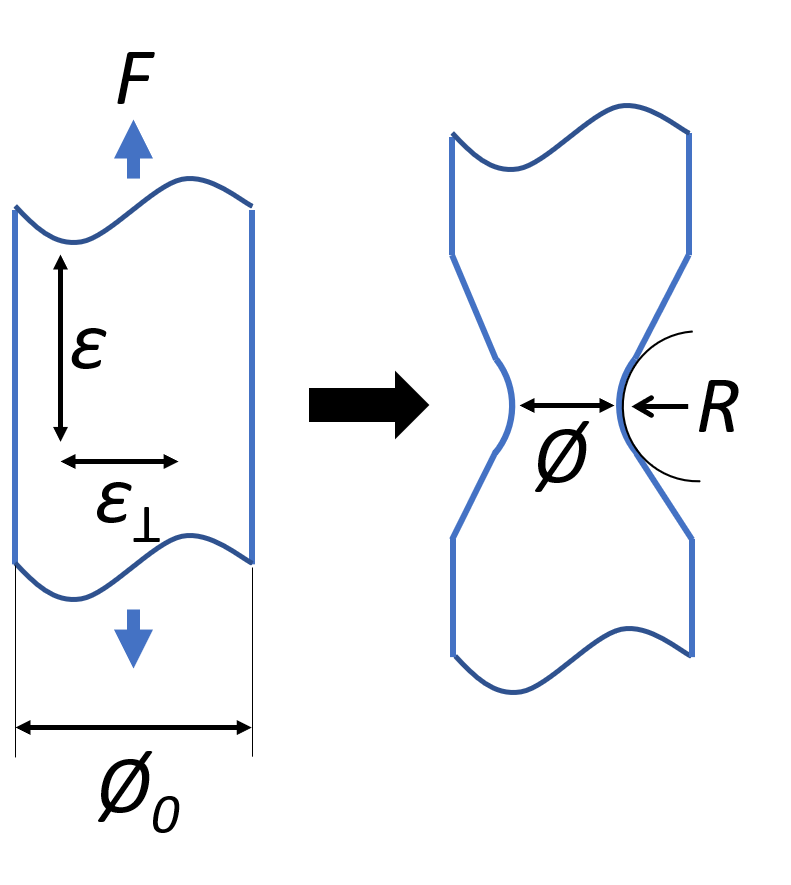}
	\caption{A cylindrical section is elongated due to the action of a force $F$. The longitudinal strain in the direction of the cylinder axis is denoted as $\varepsilon$, and the perpendicular strain is denoted as $\varepsilon_{\bot}$. The deformation leads to a diameter reduction from $\phi_0$ to $\phi$. For sufficiently large strains, necking with a local radius $R$ occurs.}
	\label{fig:cylinder_strain}
\end{figure}

The definition of true strain is
\begin{equation}
	\varepsilon = \mathrm{ln}\frac{L}{L_0}
\end{equation}
$L$ and $L_0$ are the current and initial length of a sample gauge region.
We now restrict ourselves to axis-symmetric specimens only. In this case, the longitudinal strain may be
calculated from the diameter change of the specimen. Referring to
Fig.~\ref{fig:cylinder_strain}, and starting from the definition of Poissons's ratio $\nu$, we have:

\begin{eqnarray}
	\nu & = & -\frac{\varepsilon}{\varepsilon_{\bot}} \\
	\nu & = & -\frac{\varepsilon}{\mathrm{ln}\frac{\phi}{\phi_0}} \\
	\varepsilon & = & \frac{1}{\nu} \; \mathrm{ln}\frac{\phi_0}{\phi} \label{eq:eps_from_dia}
\end{eqnarray}
Note that, in this context, $\nu$ does not refer only to the elastic Poisson's ratio at infinitesimally small strains, but to the ratio of transverse to longitudinal strains for any deformation. $\nu$ must be known before Eq.~\ref{eq:eps_from_dia} can be invoked. For many metals, however, plastic yielding is known to be an isochoric process with $\nu=0.5$, and we will assume in the following that $\nu$ is a known quantity. We will show below how the instantaneous diameter $\phi$ can be accurately identified from optical recordings, thus facilitating strain measurement.

With the instantaneous diameter known, the cross-section of axis-symmetric specimens is also known. Thus, the average stress in the specimen can be calculated with the instantaneous force,
\begin{equation}
	\sigma_{avg} = \frac{F}{A} \label{eq:sigma_avg}
\end{equation}
Note, that $\sigma_{avg}$ is, in general, not equivalent to the 1D uniaxial true
stress. This is only the case for a homogenous uniaxial deformation. If necking
is present, additional radial and hoop stresses exist, which lead to an increase
in the measured force compared to the case without necking. Thus,
Eq.~\ref{eq:sigma_avg} will overestimate the uniaxial strength of a material
when necking is present. This phenomenon, and a correction approach, was first
described by P.W. Bridgman in 1952 \cite{bridgmanStudiesLargePlastic1952}. He
developed an approximate solution involving the current diameter $\phi$ and the
radius of curvature $R$ of the neck, allowing to recover the uniaxial
stress from the average stress:

\begin{equation}
	\sigma \approxeq \sigma_{avg} \times \left[ \left( 1 + \frac{4R}{\phi} \right) \mathrm{ln}\left( 1 + \frac{\phi}{4R} \right) \right]^{-1} \label{eq:bridgman_correction} 
\end{equation}

Numerous articles have investigated the validity of Bridgman's assumptions and
accuracy of his correction method, see e.g. \cite{choungStudyTrueStress2008,
kweonDeterminationTrueStressstrain2020, lingUniaxialTrueStressStrain1996, murataStressCorrectionMethod2018,
wangExperimentalNumericalCombinedMethod2016}. A recent review \cite{tuStressStrainCurves2020}
summarizes these works. The consensus seems to be that Bridgman's correction
produces errors less than 5\% if both $\phi$ and $R$ are accurately known. Even
better corrections are available, but these typically involve additional
information about the material, e.g. the hardening exponent of the plastic yield
curve, or additional numerical simulations. For the rest of this paper, we will
use Bridgman's original correction.

Bridgman also developed an expression for the stress triaxiality based on the ratio of $\phi$ and $R$. This quantity is important for failure models such as the Johnson-Cook model, where failure strain depends on the stress triaxiality. Bridgman's original solution is
\begin{equation}
	\eta \approxeq \frac{1}{3} + \mathrm{ln}\left( 1 + \frac{\phi}{4R} \right)
	\label{eq:bridgman_eta} 
\end{equation}
An improved correction method was found by Bai \textit{et al.} \cite{baiApplicationStressTriaxiality2009} for the case of cylindrical specimens. This correction differs by an additional factor of $\sqrt{2}$, i.e.,
\begin{equation}
	\eta \approxeq \frac{1}{3} + \sqrt{2} \mathrm{ln}\left( 1 + \frac{\phi}{4R} \right)
	\label{eq:bai_eta} 
\end{equation}
In this work we use the expression given by Bai \textit{et al.}.

\subsection{Contour imaging}

\begin{figure}
	\centering
	\includegraphics[width=120mm]{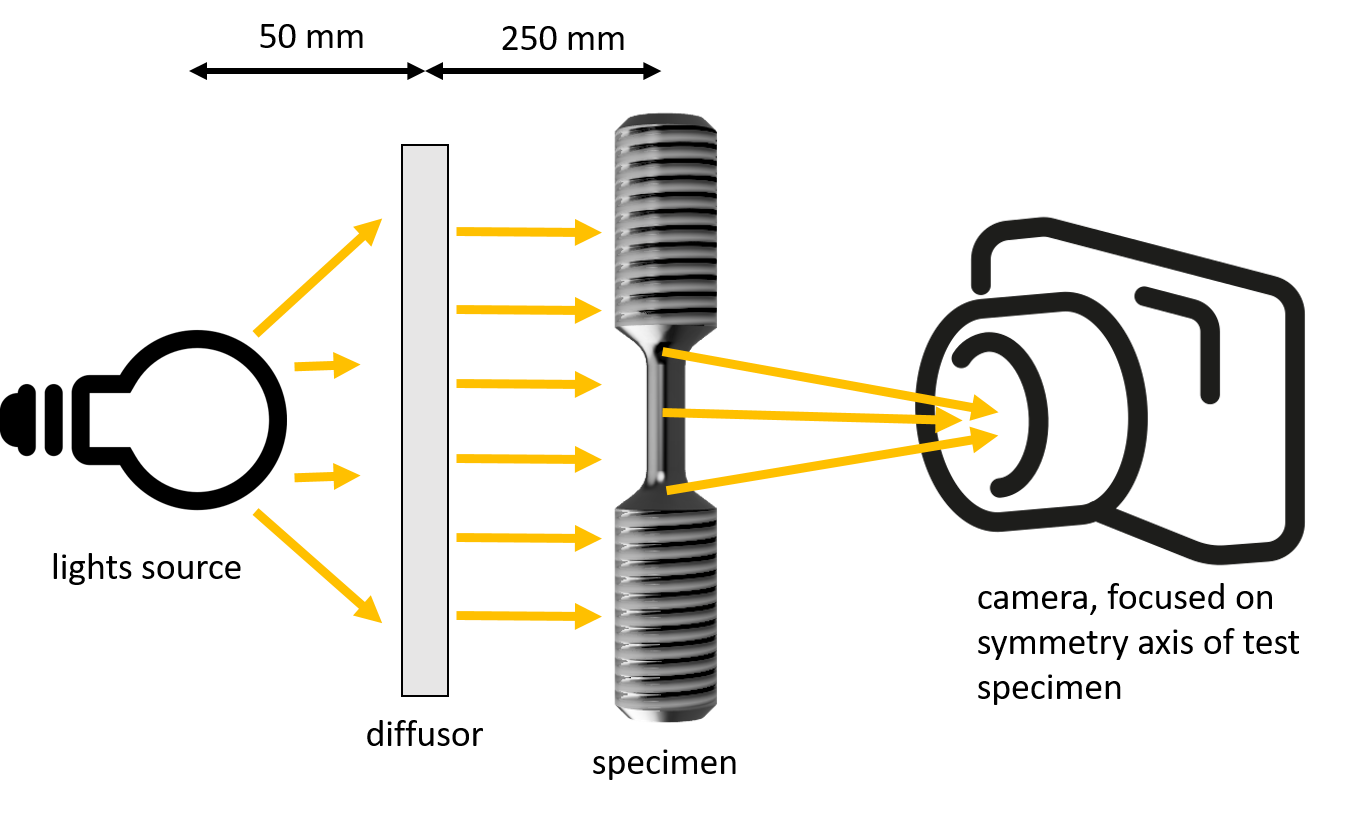}
	\caption{
		Setup of backlight contour imaging. A concentrated light source is diffused using a semi-transparent screen. The shadow image of a tensile test is recorded using a conventional camera. The focal plane of the camera is set to the	symmetry axis of the cylindrical specimen.}
	\label{fig:constrain_setup}
\end{figure}

To measure the specimen diameter and, eventually, the ensuing neck radius during
a tensile test, a backlight technique is employed in this work.
Fig.~\ref{fig:constrain_setup} shows the experimental setup. A light source is
placed in front of a diffusing screen, e.g. semi-transparent plastic. A camera
records shadow images of the specimen. It is important to set the center the
focus plane onto the symmetry axis of the specimen. In this way, the radial
contour does not move out of focus when the specimen diameter reduces as the
tensile test progresses.

\subsection{Contour analysis}
Automatic extraction of the instantaneous minimum specimen diameter and the neck
curvature around the image is accomplished by a computer program. Its source
code is published online \cite{ganzenmullerConStrainProgramPerform} and we
discuss only its salient features here. The workflow is given in the following:
\begin{enumerate}
	\item Load an ordered sequence of images and iterate over the set.
	\item Binarize each image to black and white pixels only using Otsu's adaptive thresholding method \cite{otsuThresholdSelectionMethod1979} which requires no predefined threshold.
	\item Obtain the specimen's contour from the contrast transition between white background and black specimen. For this, the Canny edge detector \cite{cannyComputationalApproachEdge1986} is used with a Gaussian filter kernel of width $\phi_0/100$ px. The results of this operation are two contour lines.
	\item Search for the location $x_{min}$ where the distance between the two contours is minimal. Store the current minimum diameter $\phi$.
	\item For each contour line, determine the neck radius $R$. To this end, The neck region is identified in relation to the diameter: a subset of the contour, centered on $x_{min}$ and of length $\phi/2$ px is isolated. A circle is fit to this subset using Kanatani's hyper-accurate method \cite{kanataniEllipseFittingHyperaccuracy2006}, which has been shown to compare favourably to other fitting methods \cite{al-sharadqahErrorAnalysisCircle2009}.
	\item With $\phi$ and $R$ known, compute the true strain and the triaxiality factor according to Eqns.~\ref{eq:eps_from_dia} and \ref{eq:bai_eta}. Also compute the current cross section $A=\pi \phi^2 /4$, the average true stress ,and the corrected average stress according to Bridgman using Eqns.~\ref{eq:sigma_avg} and \ref{eq:bridgman_correction}.
\end{enumerate}

Note, that only two user-set parameters are required for this algorithm: The
Canny edge kernel width and the extent of the necking region. Both are defined
in relation to the current minimum specimen diameter and measured in pixel
units. With the choices shown above, the algorithm produces typical strain
errors smaller than 1\% if the diameter $\phi$ is resolved with at least 200 px.

%==============================================================================
%==============================================================================
%==============================================================================

\section{Validation with simulation data}
\begin{figure}
	\centering
	\includegraphics[width=6cm]{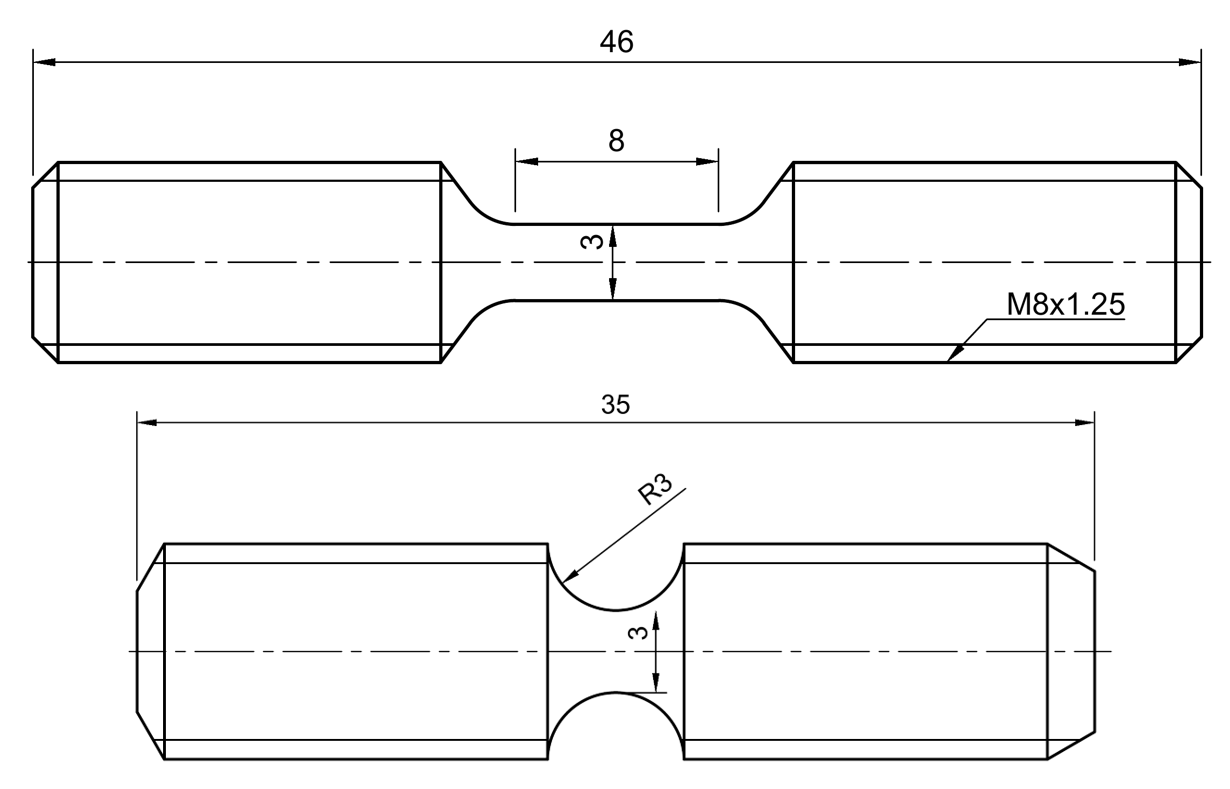}
	\caption{Sketch of the parallel and notched specimen geometries used in thios work. All dimensions are in mm.}
	\label{fig:specimens}
\end{figure}

\begin{figure}
	\centering
	\begin{tabular}{lr}
		\subf{\includegraphics[width=55mm]{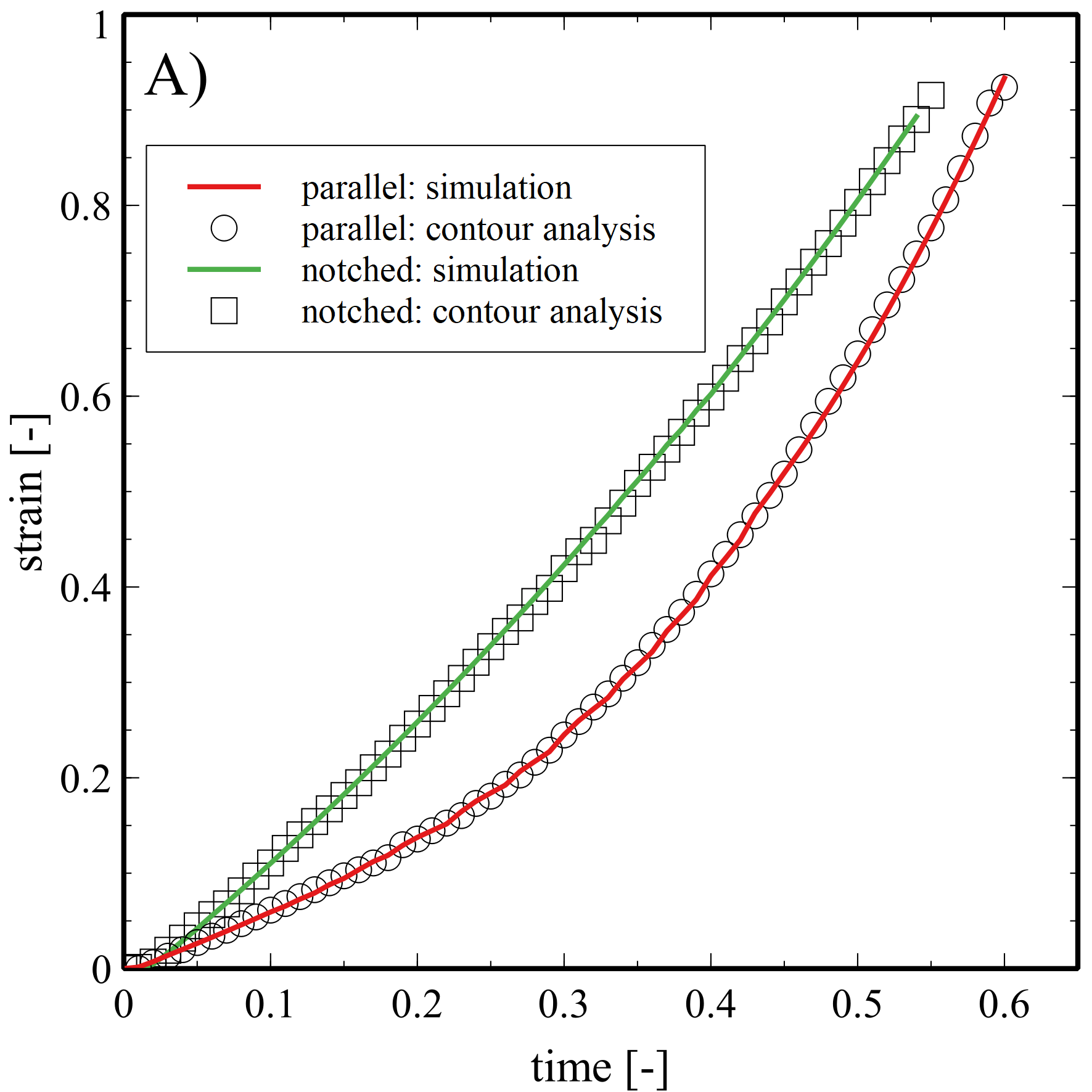}}
		{\\}
		&
		\subf{\includegraphics[width=55mm]{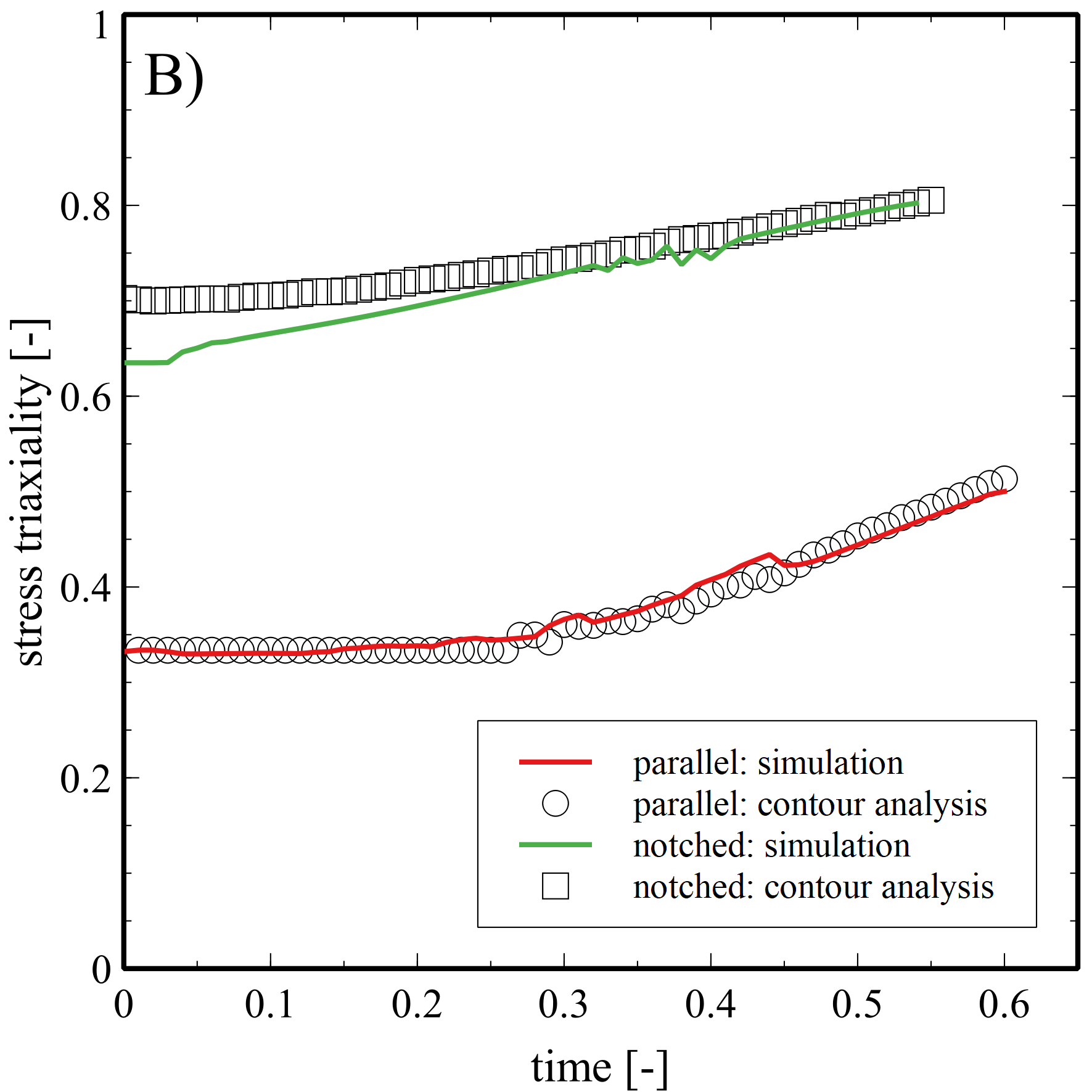}}
		{\\}
		\\
		\subf{\includegraphics[width=55mm]{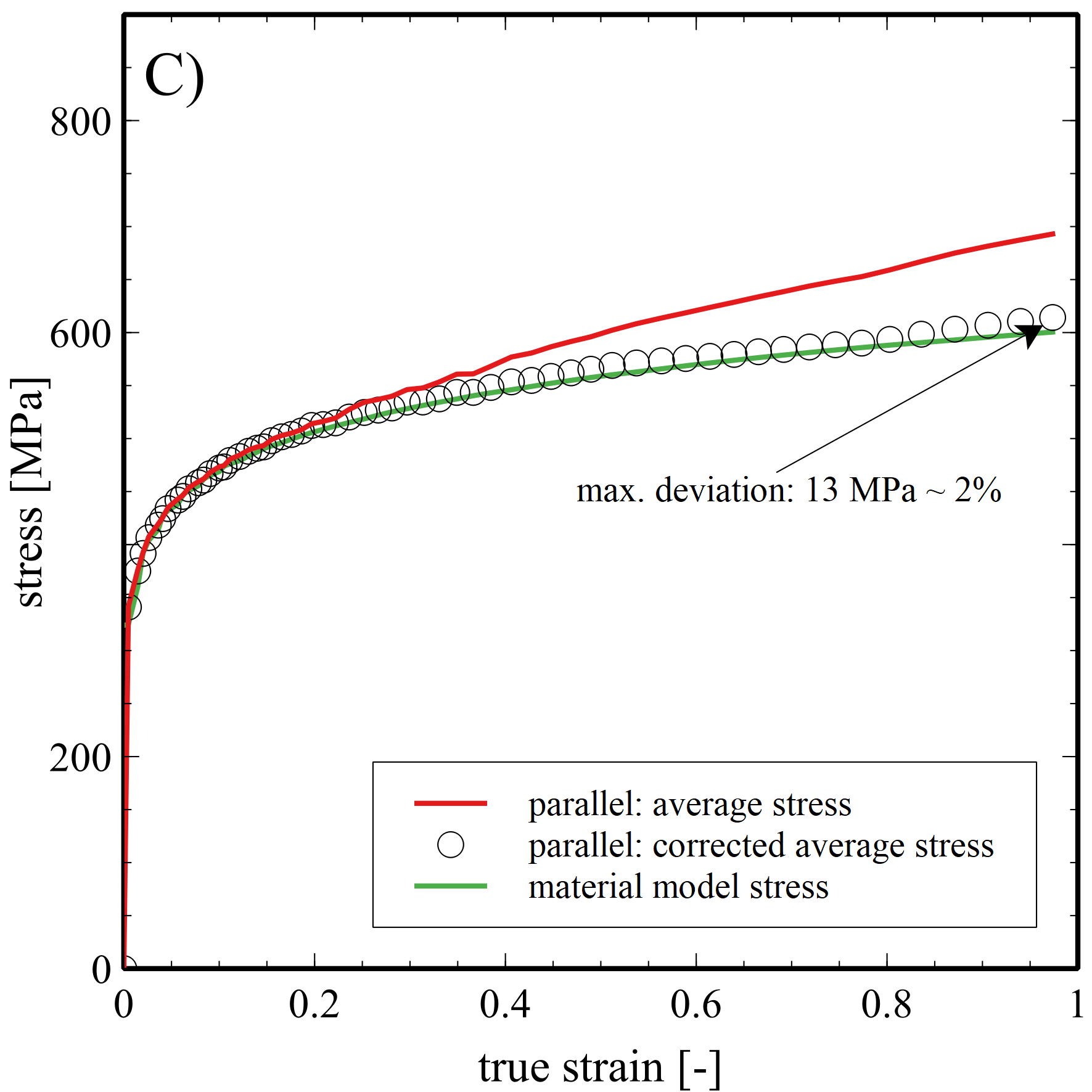}}
		{\\}
		&
		\subf{\includegraphics[width=55mm]{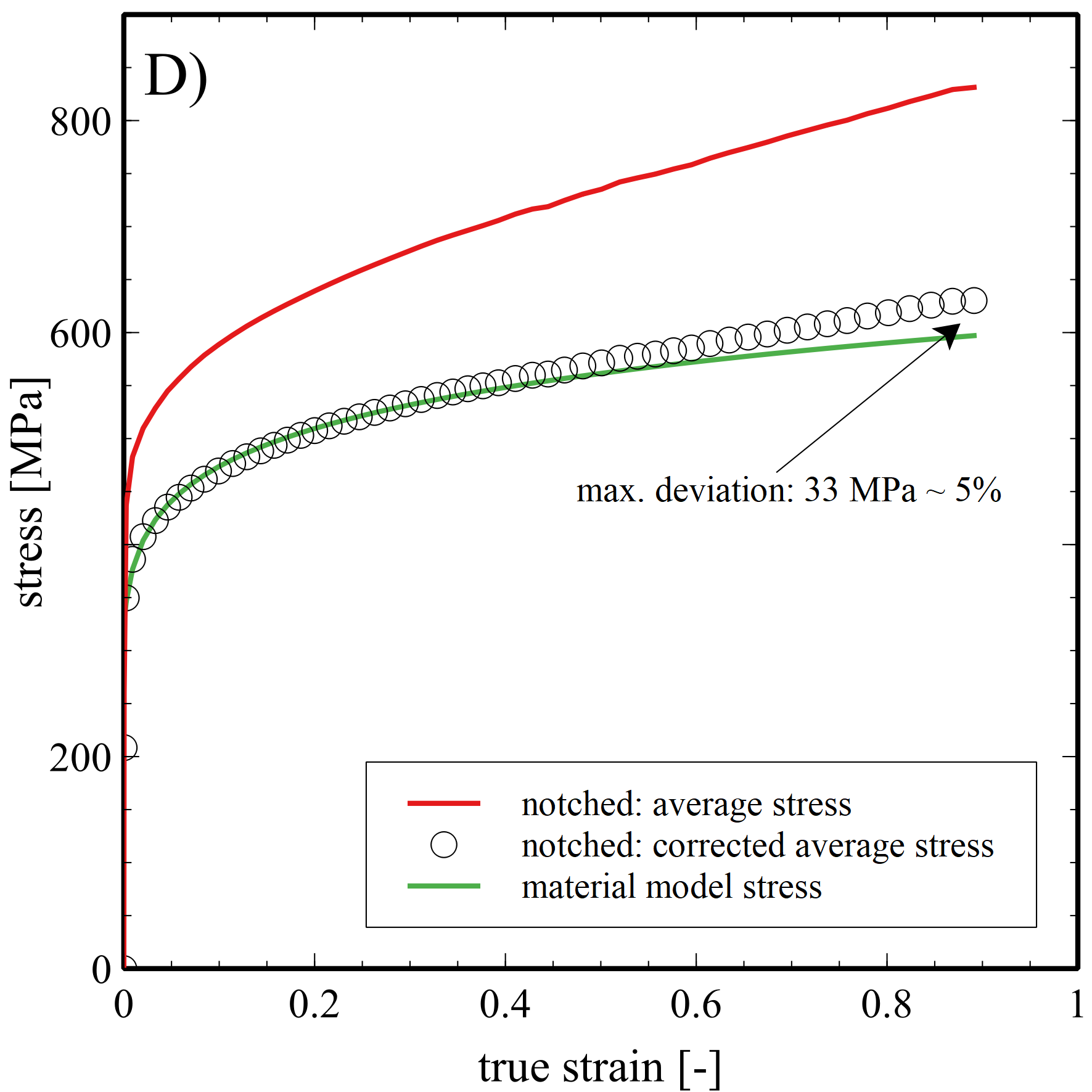}}
		{\\}
	\end{tabular}
	
	\caption{ Comparison between exact simulation data and the results of our
        contour analysis algorithm. \textbf{A$)$} compares total uniaxial strain
        for specimens with paralle and notched geometries with their simulation
        counterparts. \textbf{B$)$} compares stress triaxiality between contour
        analysis and simulation. \textbf{C$)$} and \textbf{D$)$} show true
        average stress, c.f. Eq.~\ref{eq:sigma_avg}, which overestimates the
        known material stress state if triaxiality effects are present. The same
        graphs also show the corrected true stress, obtained with
        Eq.~\ref{eq:bridgman_correction}, which agrees well with the known
        material stress. While \textbf{C$)$} displays results for parallel gauge region specimen, while \textbf{D$)$} shows results for a notched
        specimen.}
	\label{fig:simvalid}
\end{figure}

To test the accuracy of the method, synthetic image data are prepared using the
Finite Element code LS-Dyna. Two tensile specimen geometries are considered, one
with a parallel gauge region, and one with a notch, see
Fig.~\ref{fig:specimens}. The material model is chosen as elastoplastic with
kinematic hardening, Young's modulus of 70 GPa, Poisson's ratio 0.3, initial
yield stress 400 MPa and tangent modulus 100 MPa. $J_2$-plasticity with radial
return is used. An axissymmetric discretization with a fine mesh size of ~0.02
mm in the gauge region is used. Images of the simulation with white background
and all-black finite elements are output at regular intervals, serving as
synthetic experimental shadow images. These images are then analyzed using our
algorithm and the results are reported in Fig.~\ref{fig:simvalid}. For the
purpose of comparison, the simulation data are evaluated as an average over the
diameter of the gauge region, at the center of the neck.
Fig.~\ref{fig:simvalid}~\textbf{A)} shows that the total strain (elastic +
plastic) at the center of the neck is well captured by the contour analysis for
both specimen geometries. Fig.~\ref{fig:simvalid}~\textbf{B)} compares the
simulated stress triaxiality with the algorithm's prediction using
Eq.~\ref{eq:bai_eta}. The agreement is satisfactory, given the difficulty of
estimating stress triaxiality otherwise, and the approximate nature of
Eq.~\ref{eq:bai_eta}. A validation of the stress-strain relationship is
presented in Fig.~\ref{fig:simvalid}~\textbf{C)} and \textbf{D)}. Here, the
average stress is computed by our by contour analysis algorithm from the local true strain and the force, using Eq.~\ref{eq:sigma_avg}. This stress is compared to the known stress/strain relation which was input to the simulation, denoted here
as the material stress. It is evident that the stress computed in this way
agrees well at small strains, but deviates for larger strains. The reason is the
presence of stress triaxiality after the onset of necking. Correcting for this
using Eq.~\ref{eq:bridgman_correction}, the corrected true stress is obtained,
which agrees well with the known material stress: the maximum error is less than
2\% for the specimen with parallel gauge region, and less than 5\% for the
notched sample.

The comparison with simulation data shows that the contour strain evaluation
method is capable of measureing large strains with good accuracy, and estimating
stress triaxiality with a sufficient accuracy such that equivalent uniaxial
stress/strain curves can be obtained from experiment. The following section
applies the method to real experiments.

\FloatBarrier

%==============================================================================
%==============================================================================
%==============================================================================

\section{Application to experiment}
\subsection{Materials}
Specimens according to the dimensions shown in Fig.~\ref{fig:specimens} were
manufactured from stainless steel, AISI 303 (X8CrNiS18-9) using CNC turning from
10 mm bar stock. For this material, ultimate tensile strength is reported by the
manufacturer as approximately 750 MPa.

\subsection{Experiment setup}
\begin{figure}
	\centering
	\includegraphics[width=\textwidth]{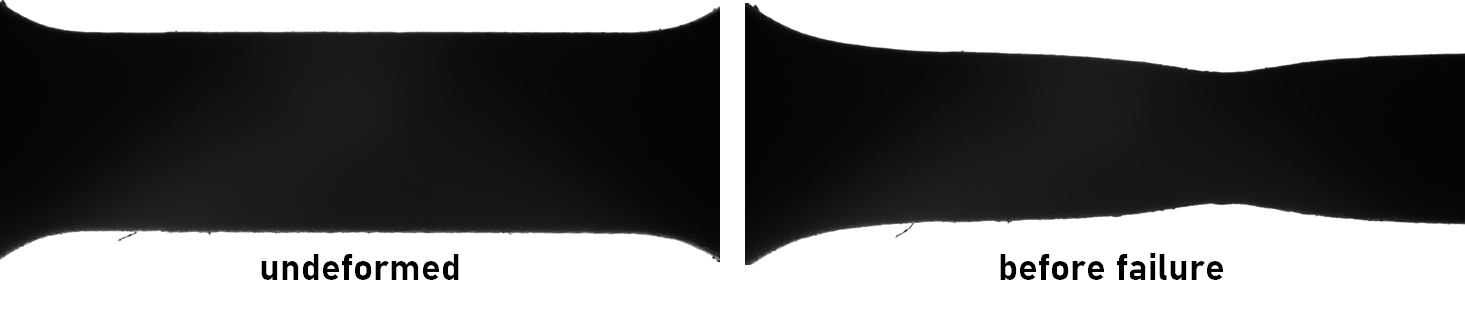}
	\caption{Unprocessed examples of acquired images for the straight specimen type considered in this work.}
	\label{fig:exp_img}
\end{figure}

A universal testing machine (Zwick-Roell Z100) with a 100 kN load-cell (accuracy
better than 0.1\% of reading for force value > 200 N) is used in displacement
control mode. The speed is set to 0.48 mm/min, corresponding to a nominal strain
rate of $10^{-3}$ /s for the specimen with a parallel gauge region of 8 mm
length. For imaging, a monochrome camera (Basler acA4112-20um) with 4096x3000 px resolution
in combination with a 50 mm lens is used, such that the initial diameter of the
specimen is resolved with $\phi_0~1000$ px. Background lighting is realised
using a 50 W COB-LED with a light-emitting area of 24x40 mm$^2$ in combination
with 3 mm thick white acrylic glass as a diffusor. Fig.~\ref{fig:exp_img} shows
represenative images acquired during the experiment.

To compare the contour strain method with the established Digital Image
Correlation (DIC) technique, a second camera with identical specifications is
used. A speckle pattern is applied using an airbrush, resulting in typical
speckle sizes of ~8 px. The commercial DIC code GOM correlate is used with the
default settings for this resolution of facet size of 51 px and point distance
20 px.

% ============================================================================
\subsection{Results}
\begin{figure}
	\centering
	\begin{tabular}{lr}
		\subf{\includegraphics[width=55mm]{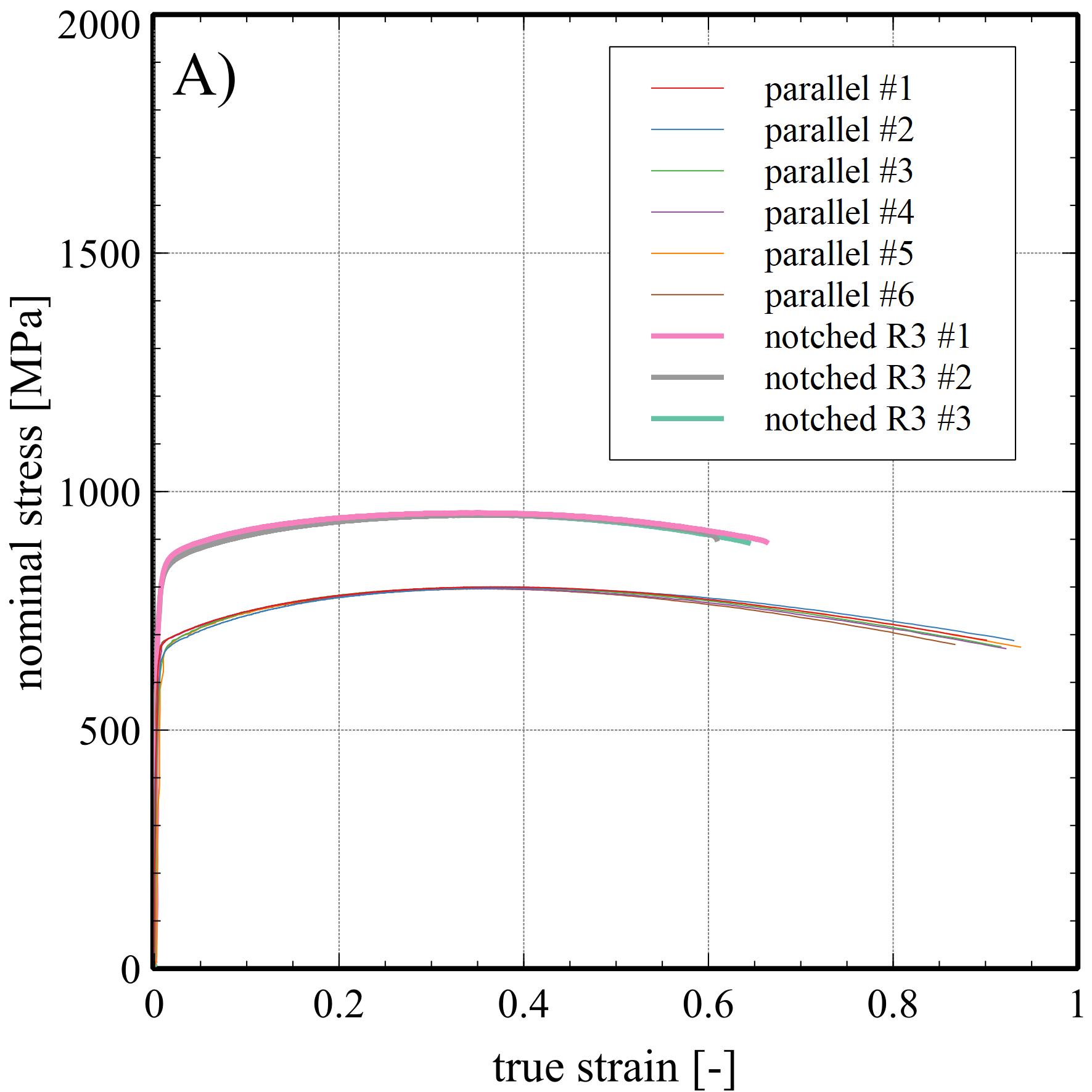}}
		{\\}
		&
		\subf{\includegraphics[width=55mm]{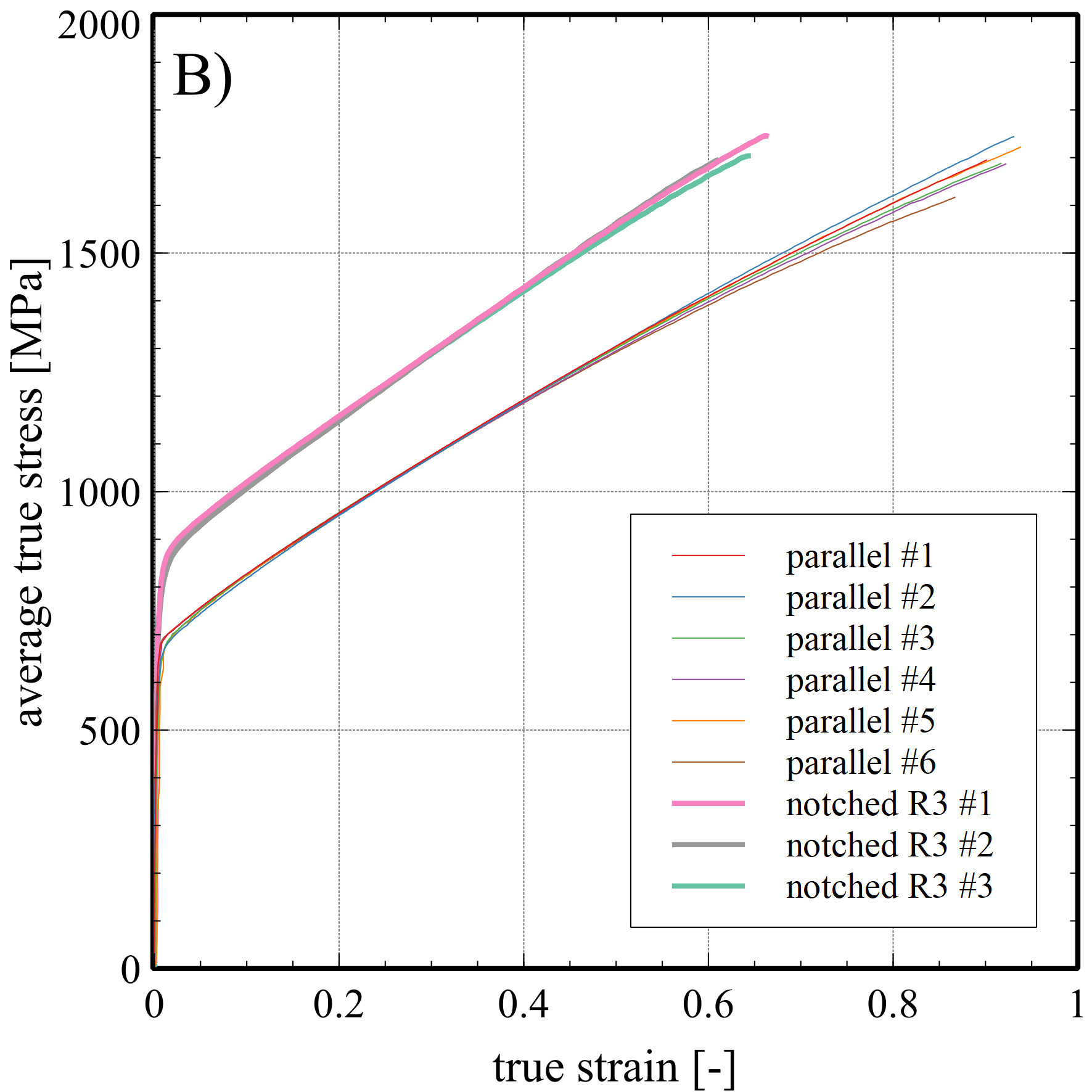}}
		{\\}
		\\
		\subf{\includegraphics[width=55mm]{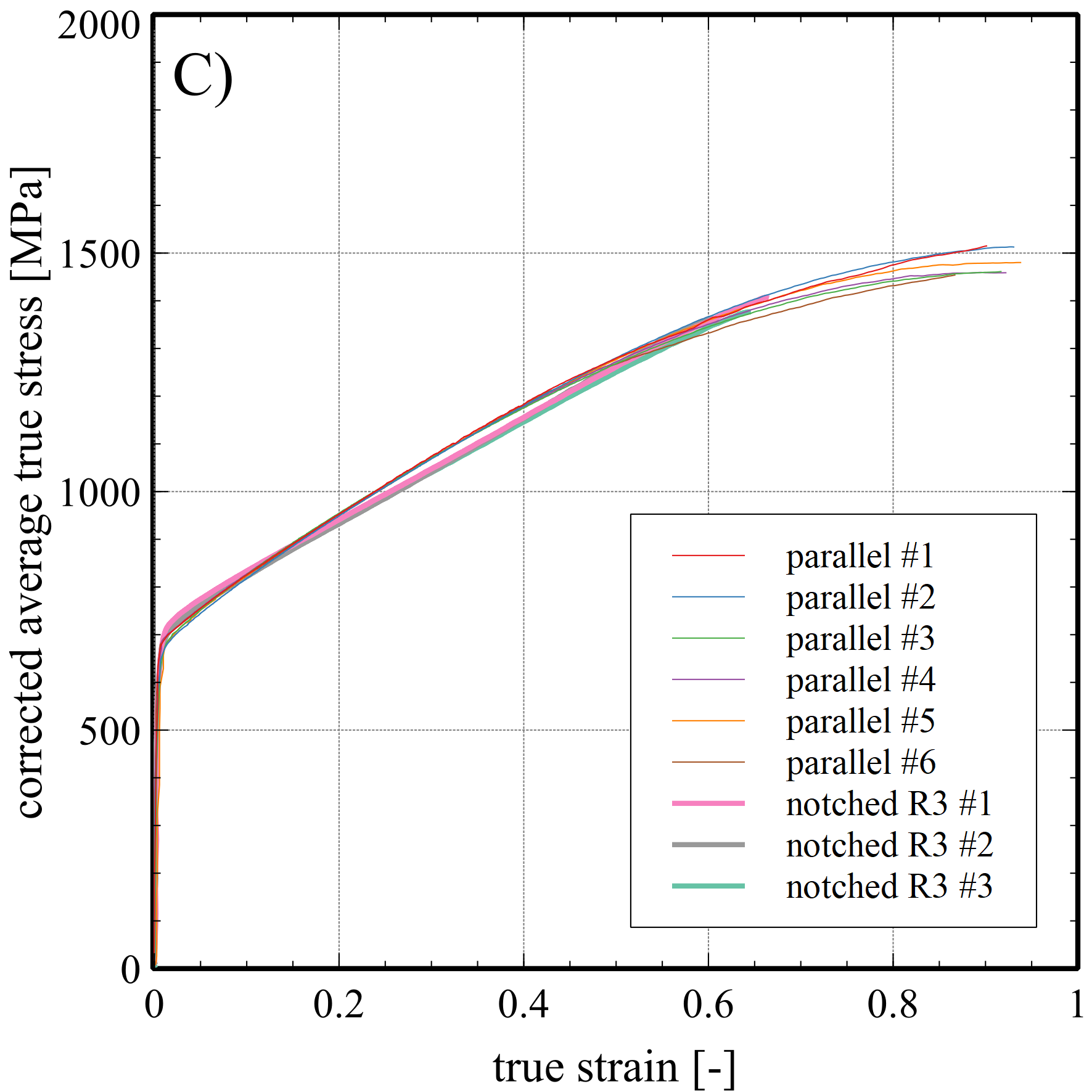}}
		{\\}
		&
		\subf{\includegraphics[width=55mm]{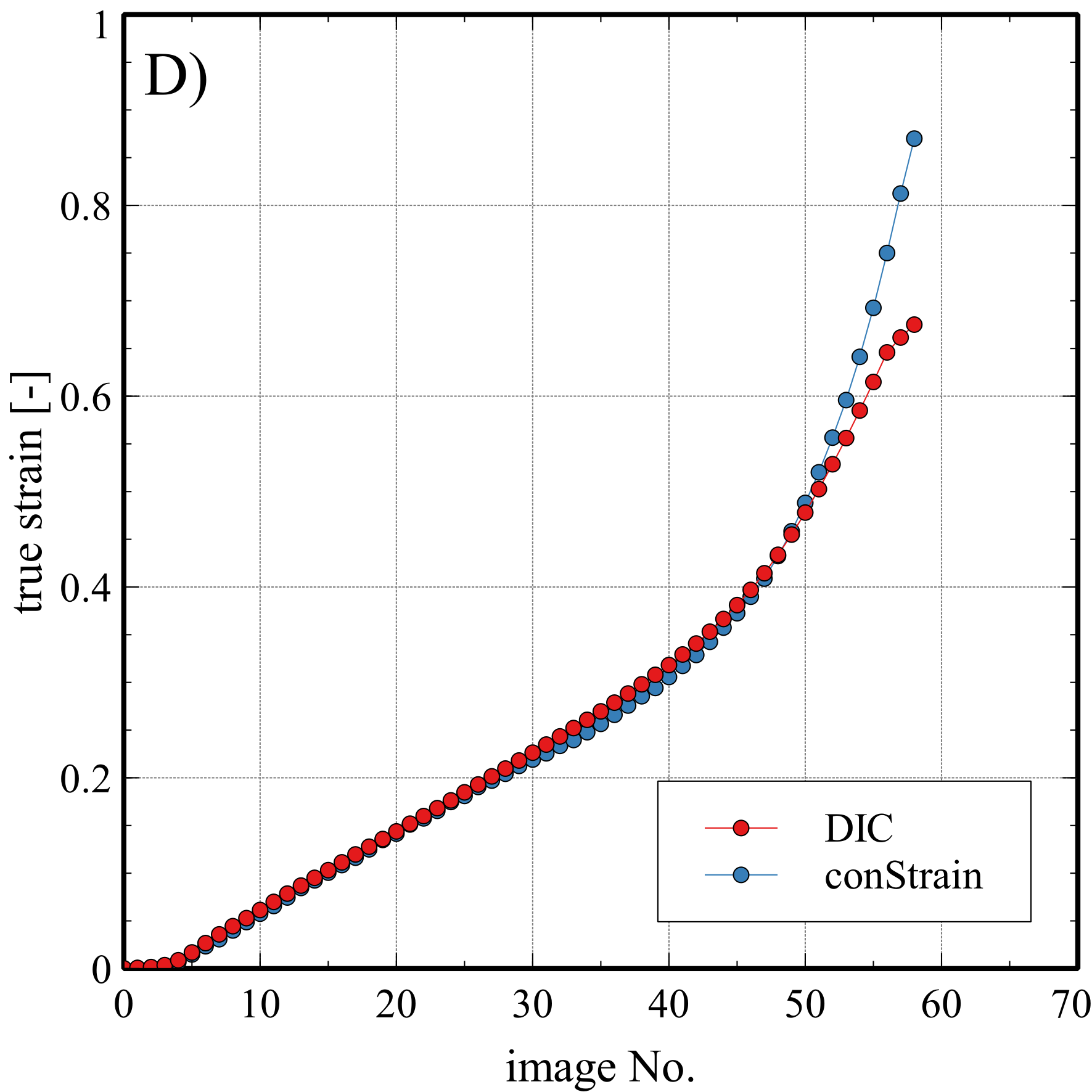}}
		{\\}
	\end{tabular}
	
	\caption{ Experimentally determined stress-strain diagrams for stainless
        steel specimens with parallel and notched gauge regions. \textbf{A$)$}
        shows nominal stress (c.f. equation nominal) over true strain (c.f.
        equation true strain). \textbf{B$)$} shows true stress (c.f. equation
        nominal) over true strain (c.f. equation true strain). \textbf{C$)$}
        shows true stress, corrected for triaxiality effects, (c.f. equation
        nominal), over true strain (c.f. equation true strain). The stress
        curves are clearly separated between notched and parallel specimen types
        in subfigures \textbf{A$)$} and \textbf{B$)$}, but collapse onto each
        other once triaxiality effects are accounted for, as seen in subfigure
        \textbf{C$)$}. For one parallel gauge section specimen, comparison with
        DIC evaluation is shown in \textbf{D$)$}.}
	\label{fig:results}
\end{figure}

Fig.~\ref{fig:results} shows stress/strain data obtained using the methods
described in this work. A total of 6 specimens with parallel gauge region 3
specimens with notched geometry were tested. Fig.~\ref{fig:results}~\textbf{A}
shows nominal stress (engineering stress, force over initial cross-section area)
over true strain, which is computed using contour analysis. The different
specimens geometries are grouped tightly, indicating good specimen quality and
repeatability of the analysis method. The curves for the notched specimens read
approximately 160 MPa higher stress than the parallel gauge type specimens. This
apparent stress enhancement is due to stress triaxiality. Average true stress
over true strain, as defined in Eq.~\ref{eq:sigma_avg}, is shown in
Fig.~\ref{fig:results}~\textbf{B}. These curves are of strictly monotonous
character. This is to be expected, as a negative slope (softening) would
indicate material instability. However, the groups for the different specimens
are still separated. Therefore, they do not yet present the underlying uniaxial
material behaviour. This is of course obvious for the notched specimen geometry,
which is affected from triaxiality effects already at small strains. The
parallel gauge section specimens, however, undergo necking at a strain of
approximately 35\%. From this point onwards, they are also affected by stress
triaxiality, and the average true stress is not a good measure for the
underlying uniaxial stress/strain curve. Fig.~\ref{fig:results}~\textbf{C} shows
that all data collapse onto a single stress/strain curve, if the corrected true
stress according to Eq.~\ref{eq:bridgman_correction} is considered, only the
failure strains remain different. This demonstrates that this correction
addresses the effects of stress triaxiality adequately. The resulting
stress/strain data now represents the equivalent stress and equivalent strain
for uniaxial loading, and may be used, e.g., as yield stress input curve for
simulation purposes.

Comparison with the established strain measurement technique Digital Image
Correlation is made in Fig.~\ref{fig:results}~\textbf{D}. The DIC evaluation
window is limited to those facets which are located within the necking region
only. At small strains, the agreement between both methods is good, but the
accuracy of DIC is presumably quite better due sub-pixel interpolation. At large
strains, close to specimen failure, strains reported by DIC are significantly
lower than the strain values computed by contour analysis. This can be
understood by the fact, that the regions used in DIC for strain evaluation
cannot be made arbitrarily small. Thus, DIC values are smoothed in space.
Additionally, the DIC speckle pattern deteriorates for large strains,  DIC
accuracy becomes questionable. This is of course strongly dependent on the
magnitude of failure strain and the quality of the speckle pattern and optical
recordings. Nevertheless, contour analysis is unaffected by these difficulties.

\section{Discussion and conclusion}
This work revisits the idea of G'Sell \textit{et al.}
\cite{gsellVideocontrolledTensileTesting1992}, which is to compute true strain
for axis-symmetric tensile specimens using optical contour analysis methods. In
our interpretation of this, we use image processing techniques including the
Canny-Edge filter \cite{cannyComputationalApproachEdge1986} for contrast
enhancement and a modern robust curvature fitting routine
\cite{kanataniEllipseFittingHyperaccuracy2006}. This allows us to not only
measure true strain, but also estimate stress triaxiality. There are two major
inherent limitations to this approach. It is only applicable to axis-symmetric
specimens, and ratio of transversal strain to longitudinal strain must be known
beforehand. Nevertheless, for ductile metals, this does not impose drastic
limitations, as these typically obey isochoric during plastic deformation, and
cylindrical specimen geometries are often used. Under these circumstances, the
method provides a very convenient, and accurate measurement of the equivalent
uniaxial true stress / true strain relationship as needed for constitutive
modelling. This work shows, by comparison with manufactured solutions using FEM
simulations, and application to real tensile tests, that this approach is
practical and viable. The method requires only a single camera, as the contour
does not move out of the focal plane. No 3D effects must be considered, as would
otherwise be the case with the established DIC method
\cite{weidnerReviewStrainLocalization2021}. Furthermore, no speckle pattern is
required, as the method used a shadow image obtained with background
illuminations. This means that contour strain analysis could be applied to
extreme scenarios, such as high-temperature tensile tests, where a speckle
pattern would thermally deteriorate. Our contour strain analysis technique
provides a local strain measurement directly at the minimum diameter of a
specimen. It is applicable to strong necking in ductile metals, where it also
provides an estimate of the stress triaxiality. We believe that the technique is
useful for experimental analysis and provide an opens source implementation of
our program, termed \textit{conStrain} \cite{ganzenmullerConStrainProgramPerform}.

\section*{Conflicts of Interest}
The authors declare no competing interests.

%\printendnotes

\FloatBarrier

% Submissions are not required to reflect the precise reference formatting of the journal (use of italics, bold etc.), however it is important that all key elements of each reference are included.
\bibliography{constrain}

%  \begin{biography}[PINK_PANTHER]{A.~One}
% % Optional: please include a photograph and a brief biography (up to 75 words) for each author.
%  \bigskip
%  \bigskip
%  \bigskip
% \end{biography}

\end{document}